\documentclass[prb, twocolumn]{revtex4-1} 
\usepackage{color}
\usepackage{verbatim}
\usepackage{bm}
\usepackage{amsmath}
\usepackage{amssymb}
\usepackage{graphicx}
\usepackage{esint}

\begin{document}

\title{The Coriolis field}

\author{L.~Filipe Costa}
\email{lfpocosta@math.ist.utl.pt}
\affiliation{Center for Mathematical Analysis, Geometry and Dynamical Systems, Instituto Superior T\'ecnico, Universidade de Lisboa, Portugal}

\author{Jos\'e Nat\'ario}
\email{jnatar@math.ist.utl.pt}
\affiliation{Center for Mathematical Analysis, Geometry and Dynamical Systems, Instituto Superior T\'ecnico, Universidade de Lisboa, Portugal}

\date{\today}

\begin{abstract}
We present a pedagogical discussion of the Coriolis field, emphasizing its not-so-well-understood aspects. 
We show that this field satisfies the field equations of the so-called Newton-Cartan theory, 
a generalization of Newtonian gravity that is covariant under changes of arbitrarily rotating and accelerated frames. Examples of solutions of this theory are
given, including the Newtonian analogue of the G\"odel universe. We discuss how to detect the Coriolis field by its effect on gyroscopes, of which the gyrocompass is an example.
Finally, using a similar framework, we discuss the Coriolis field generated by mass currents in general relativity, and its measurement by the Gravity Probe B and LAGEOS/LARES experiments.
\end{abstract}
\maketitle

\section{\textit{Red Planet} and the principle of equivalence}

\begin{figure}[h!]
\includegraphics[width=1\columnwidth]{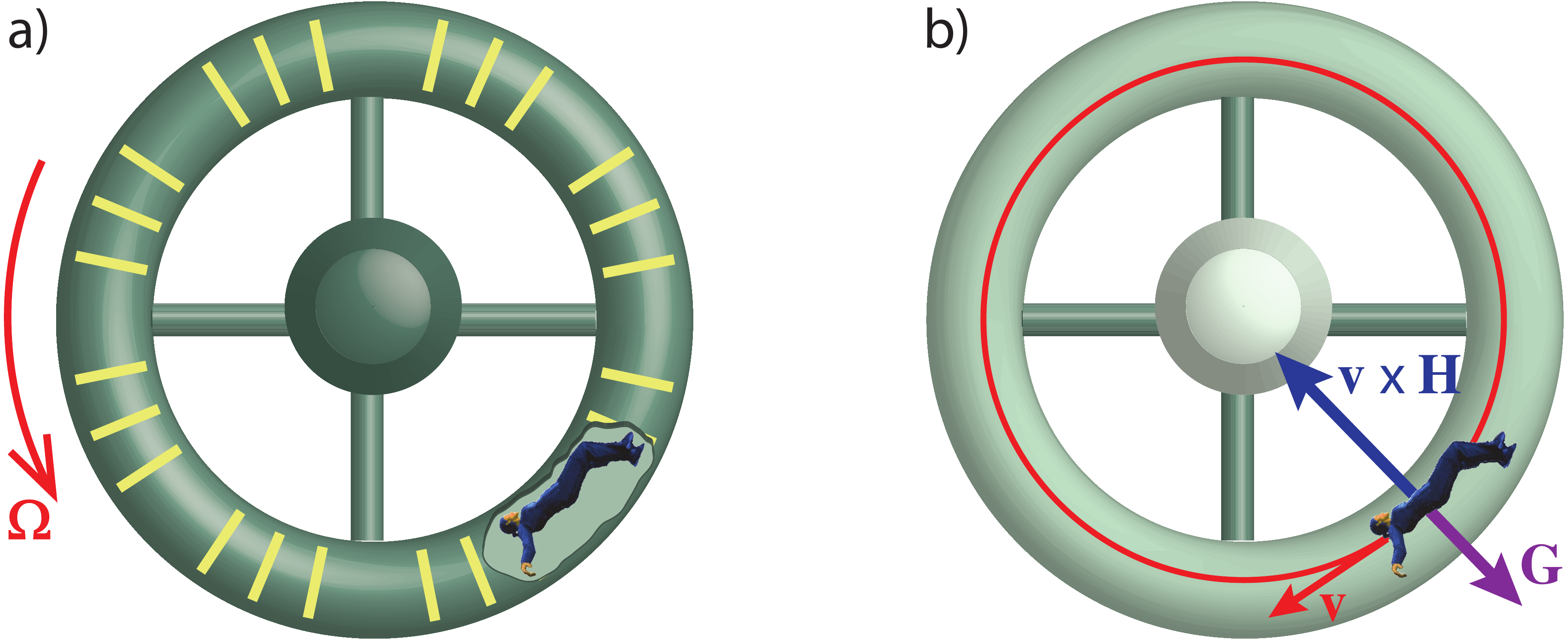} 
\protect\protect\protect\caption{Commander Bowman floating inside the rotating wheel as seen: (a) in an inertial frame;
(b) in a frame co-rotating with the wheel. From the point of view of the
inertial frame, she is freely falling and remains stationary. From
the point of view of the wheel's frame, she is in a circular orbit
under the action of two forces: a centrifugal force $m{\bf G}$ (that
pushes her outwards), and a centripetal Coriolis force $m{\bf v}\times{\bf H}=-2m{\bf G}$,
with twice the magnitude of the centrifugal force (that pushes her
inwards).}
\label{Costa_NatarioFig01} 
\end{figure}

In the movie \textit{Red Planet} (2000), a spaceship traveling to Mars
simulates gravity by using a spinning wheel.\cite{wheel} The astronauts
living inside the wheel are pushed against the outer wall, which
they perceive as the floor, by the centrifugal force.
At some point during the movie a power failure causes the wheel to
stop spinning, leading to a zero gravity environment. This event provides
the necessary dramatic setting for commander Kate Bowman to save the
day, including explosively decompressing the wheel to extinguish a
zero-$g$ fire. When she gets the main power back online the wheel starts
spinning again, and everything---including her---immediately falls to the floor.
But as noted on the popular website ``Bad Astronomy,''\cite{BA} this
is not what would happen in such a situation. Instead, imagine commander
Bowman floating motionless (with respect to some inertial frame, say the rest frame of the distant stars)
inside the wheel (see Fig.~\ref{Costa_NatarioFig01}) and assume,
for simplicity, that there is no air inside the spaceship. From the point of view
of the star fixed inertial frame it is clear that, no matter how fast the wheel
spins, she will feel no force and will continue floating motionless inside
the wheel.

If we take the perspective of the wheel's (rotating) frame and
interpret the centrifugal (inertial) force as a gravitational field
the situation looks very strange: commander Bowman is subject to a
gravitational force and yet she does not fall. After a moment's thought,
one realizes that such a thing happens all the time in ``real''
gravitational fields: it is called being in orbit. In the wheel's
rotating frame, Commander Bowman is moving along a circle with the
right velocity so that she does not fall. However, the gravitational
field (centrifugal force) points outward, away from the center of
her trajectory. So how can she be in orbit in this gravitational
field? Something is surely missing.

\section{Non-relativistic inertial forces: Newton-Cartan theory\label{sec:Non-relativistic-inertial-forces}}

The missing ingredient is, of course, the Coriolis force; it is this
velocity-dependent force that pushes commander Bowman towards the
center of her orbit. This reminds us that Einstein's equivalence principle,\cite{Einstein}
when applied to arbitrarily accelerated and rotating frames, requires
the introduction of a magnetic-like Coriolis field.

In this section we will revisit the problem of the inertial forces
arising in an arbitrarily accelerated and rotating (rigid) frame,
and obtain the so-called Newton-Cartan equations, which will be useful in our
further study of the Coriolis field in the next sections.
Let ${\bf x}(t)$ be the position vector of a particle of mass $m$
in an inertial frame $S$ (say, the rest frame of the distant stars),
and let ${\bf r}(t)$ be the position vector of the same particle
in an accelerated and rotating frame $S'$ (for example, the wheel's
frame). If we denote the position of the origin of $S'$ in $S$ by
${\bf x}_{0}(t)$ then 
\begin{equation}
{\bf x}(t)=R(t){\bf r}(t)+{\bf x}_{0}(t)\ ,\label{eq:x-r}
\end{equation}
for some time-dependent rotation matrix $R(t)$. In the example above,
for instance, ${\bf x}_0(t)$ represents the position of the center of the wheel and
\begin{equation}
R(t) =
\left(
\begin{matrix}
\cos(\Omega t) & - \sin(\Omega t) & 0 \\
\sin(\Omega t) & \cos(\Omega t) & 0 \\
0 & 0 & 1
\end{matrix}
\right),
\label{eq:R}
\end{equation}
where we have assumed that the wheel is rotating around the $z$-axis with
constant angular velocity $\Omega$.

If we assume that
the particle is subject to a gravitational field ${\bf g}$ plus a
nongravitational force $\mathbf{f}$, then the particle's equation
of motion will be 
\begin{equation}
m\ddot{{\bf x}}=m{\bf g}+\mathbf{f}\ ,\label{motionS}
\end{equation}
where dots represent time derivatives.
As usual, we take the gravitational field to satisfy 
\begin{equation}
\nabla\times{\bf g}={\bf 0},\qquad\nabla\cdot{\bf g}=-4\pi G\rho\ ,
\end{equation}
where $G$ is Newton's constant and $\rho$ is the mass density of
the matter generating the field.

To write the equation of motion in the frame $S'$, we note that 
\begin{equation}
\dot{{\bf x}}=R\dot{{\bf r}}+\dot{R}{\bf r}+\dot{{\bf x}}_{0}\ . \label{velocity0}
\end{equation}
Since $R$ is a rotation matrix, its inverse coincides with its transpose:
$R^{T}=R^{-1}$. Hence 
\begin{equation}
R^{T}R=I\Rightarrow\dot{R}^{T}R+R^{T}\dot{R}=0\Leftrightarrow R^{T}\dot{R}=-\left(R^{T}\dot{R}\right)^{T},
\end{equation}
that is, the matrix $A=R^{T}\dot{R}$ is antisymmetric (and thus it has
only three independent components, say $\Omega^1$, $\Omega^2$ and $\Omega^3$). Consequently, we can write $A_{ij}=\epsilon_{ikj}\Omega^{k}$,
where $\epsilon_{ijk}$ is the Levi-Civita alternating symbol, 
which means $A{\bf r}={\bf \Omega}\times{\bf r}$ and thus
\begin{equation}
\dot{R}{\bf r}=RA{\bf r}=R({\bf \Omega}\times{\bf r})\ . \label{seis}
\end{equation}
Note from Eqs.~\eqref{velocity0} and \eqref{seis} that the velocity vector of a particle at rest in $S'$ is
\begin{equation} \label{velocityrest}
\dot{{\bf x}}=R({\bf \Omega}\times{\bf r})+\dot{{\bf x}}_{0} = (R{\bf \Omega})\times(R{\bf r})+\dot{{\bf x}}_{0} \ ,
\end{equation}
where in the last equality we used the fact that rotations preserve cross products, since they preserve lengths, angles, and orientations.
This equation tells us that ${\bf \Omega}$ is the angular velocity of $S'$ with
respect to $S$, expressed in $S'$ (that is, the angular velocity
in $S$ is $R{\bf \Omega}$). Note that since $R=R(t)$ depends only on time, then so does ${\bf \Omega}={\bf \Omega}(t)$.

For an arbitrarily moving particle we have, from Eqs.~\eqref{velocity0} and \eqref{seis},
\begin{equation}
\dot{{\bf x}}=R\dot{{\bf r}}+R({\bf \Omega}\times{\bf r})+\dot{{\bf x}}_{0}\label{velocity}
\end{equation}
and 
\begin{equation}
\ddot{{\bf x}}=R\ddot{{\bf r}}+2R({\bf \Omega}\times\dot{{\bf r}})+R\left[{\bf \Omega}\times({\bf \Omega}\times{\bf r})\right]+R(\dot{{\bf \Omega}}\times{\bf r})+\ddot{{\bf x}}_{0}\ .
\end{equation}
The equation of motion in $S'$ is then 
\begin{equation}
m\ddot{{\bf r}}=m{\bf G}+m\dot{{\bf r}}\times{\bf H}+{\bf F}\ ,\label{eq:AccelS'Cartesian}
\end{equation}
where ${\bf F}=R^{-1}{\bf f}$ is the nongravitational force expressed
in $S'$, and we have defined the field 
\begin{equation}
{\bf G}=R^{-1}({\bf g}-\ddot{{\bf x}}_{0})-{\bf \Omega}\times({\bf \Omega}\times{\bf r})-\dot{{\bf \Omega}}\times{\bf r}\ ,\label{G}
\end{equation}
and the {\em Coriolis field} 
\begin{equation}
{\bf H}=2{\bf \Omega}\ .\label{H}
\end{equation}
In other words, to explain the motion of the particle in the frame
$S'$ we need two fields: a 
field ${\bf G}$, consisting of the preexisting gravitational field
plus inertial terms,\cite{inertialterms} and a Coriolis field ${\bf H}$,
which is twice the angular velocity of $S'$ and which gives rise to the
magnetic-like velocity-dependent Coriolis force $m\dot{{\bf r}}\times{\bf H}$.

Note from Eq.~\eqref{eq:AccelS'Cartesian} that ${\bf G}$ is minus
the nongravitational force per unit mass acting on observers at rest
in $S'$; this is exactly what a weighing scale placed in $S'$ will
measure. Moreover, Eq.~\eqref{velocityrest} implies that observers at
rest in $S'$ are seen to move in $S$ along the velocity field 
\begin{align}
{\bf v}({\bf x},t) & =(R{\bf \Omega})\times(R{\bf r})+\dot{{\bf x}}_{0}\nonumber \\
 & =(R{\bf \Omega})\times({\bf x}-{\bf x}_{0})+\dot{{\bf x}}_{0}\ ,\label{eq:v2}
\end{align}
whose vorticity (in $S$) is 
\begin{equation}
{\boldsymbol{\omega}}\equiv\frac{1}{2}\nabla\times{\bf v}=\frac{1}{2}\nabla\times\left[(R{\bf \Omega})\times{\bf x}\right]=R{\bf \Omega}\ .\label{eq:VorticityNewtonian}
\end{equation}
Here we have used the vector identity 
\begin{equation}
\nabla\times(\mathbf{a}\times\mathbf{b})=\mathbf{a}(\nabla\cdot\mathbf{b})-(\mathbf{a}\cdot\nabla)\mathbf{b}\ ,\label{eq:VectorId}
\end{equation}
which holds for a spatially constant vector $\mathbf{a}$,
together with the identities $\nabla \cdot {\bf x} = 3$ and $(\mathbf{a}\cdot\nabla)\mathbf{x} = \mathbf{a}$.
Thus, we see that ${\bf H}=2{\bf \Omega}=R^{-1}(2{\boldsymbol{\omega}})$
is twice the vorticity of the observers at rest in $S'$ (as seen in $S$ and expressed
in $S'$). These observations will be important for the comparison
with the relativistic inertial forces in Sec.~\ref{sec:Inertial-forces-in_GR}.

Since the Coriolis field does not depend on the space coordinates, as ${\bf H} = 2{\bf \Omega}(t)$, it trivially satisfies 
\begin{equation}
\nabla\times{\bf H}={\bf 0}\ ,\qquad\nabla\cdot{\bf H}=0\ .\label{eqH}
\end{equation}
Conversely, if ${\bf H}$ satisfies these equations and is spatially constant at infinity then it must be spatially constant everywhere. This fact can be seen by noting that $\nabla\times \bf{H} =0$ implies that
$\bf{H}=\nabla\psi$, for some scalar function $\psi$; then $\nabla\cdot\bf{H}=0$
implies $\nabla^{2}\psi=0$, i.e, $\bf{H}$ is the gradient of a solution of the
Laplace equation. By virtue of the Green theorem (see, e.g., Secs.~1.8-1.9 of
Ref.~\onlinecite{Jackson}), such a solution is unique (up to a constant) if boundary
conditions for its gradient are given at infinity. Therefore it must be the solution 
whose gradient, that is, ${\bf H}$, is spatially constant.

If we take the curl of Eq.~\eqref{G} we obtain 
\begin{equation}
\nabla\times{\bf G}=-2\dot{{\bf \Omega}}=-\frac{\partial{\bf H}}{\partial t}\, .
\end{equation}
To obtain this result, we: (i) used the covariance of this differential operator under rotations,\cite{Cox}
so that $\nabla\times(R^{-1}{\bf g})=R^{-1}(\nabla\times{\bf g})={\bf 0}$; (ii)
noted, from elementary vector identities, that ${\bf \Omega}\times({\bf \Omega}\times{\bf r})=({\bf \Omega}\cdot\mathbf{r}){\bf \Omega}-\Omega^{2}\mathbf{r}$
and $\nabla\times\left[({\bf \Omega}\cdot\mathbf{r}){\bf \Omega}\right]=\nabla({\bf \Omega}\cdot\mathbf{r})\times{\bf \Omega}={\bf 0}$;
and (iii) used Eq.~(\ref{eq:VectorId}) to obtain $\nabla\times(\dot{{\bf \Omega}}\times{\bf r})=2\dot{{\bf \Omega}}$.

Similarly, taking the divergence of Eq.~\eqref{G} yields 
\begin{equation}
\nabla\cdot{\bf G}=\nabla\cdot{\bf g}+2{\bf \Omega}^{2}=-4\pi G\rho+\frac{1}{2}{\bf H}^{2}\, ,\label{eqG2}
\end{equation}
where we again used the invariance of this differential operator under rotations,\cite{Cox}
$\nabla\cdot(R^{-1}{\bf g})=\nabla\cdot{\bf g}$, and noted that $\nabla\cdot(\dot{{\bf \Omega}}\times{\bf r})=0$
from the vector identity $\nabla\cdot(\mathbf{a}\times\mathbf{b})=\mathbf{b}\cdot(\nabla\times\mathbf{a})-\mathbf{a}\cdot(\nabla\times\mathbf{b})$.
Note the similarity of Eqs.~\eqref{eqH}--\eqref{eqG2} with Maxwell's
equations, apart from the absence of source and displacement currents
in the equation for $\nabla\times{\bf H}$ and the nonlinear term
in the equation for $\nabla\cdot{\bf G}$.

If there are no nongravitational forces acting on the particle, then
its mass $m$ drops out of Eq.~\eqref{eq:AccelS'Cartesian}: 
\begin{equation}
\ddot{{\bf r}}={\bf G}+\dot{{\bf r}}\times{\bf H}\ .\label{eq:AccelS'Cartesian1}
\end{equation}
This equation, together with the field equations 
\begin{equation}
\begin{cases}
{\displaystyle \nabla\times{\bf G}=-\frac{\partial{\bf H}}{\partial t}}\\
\nabla\times{\bf H}={\bf 0}\\
{\displaystyle \nabla\cdot{\bf G}=-4\pi G\rho+\frac{1}{2}{\bf H}^{2}}\\
\nabla\cdot{\bf H}=0\; ,
\end{cases} \label{NCfieldeqns}
\end{equation}
form the basis of the so-called {\em Newton-Cartan} theory of gravity.\cite{Newton-Cartan}
This theory is a generalization of the usual Newtonian gravity theory
that is covariant under changes of rigid frame, in the sense that
Eqs.~\eqref{eq:AccelS'Cartesian1} and \eqref{NCfieldeqns} are invariant
under such changes (just like the Maxwell equations and the Lorentz
force law are invariant under changes of inertial frame). In this
theory, inertial frames are thus no longer privileged---all inertial
forces are explained as a combination of the fields ${\bf G}$ and ${\bf H}$. 
Unsurprisingly, it is the (singular) limit of general relativity
as the speed of light tends to infinity.\cite{R89, D90, E91,E97}

The Newton-Cartan theory is, however, more general than the usual Newtonian theory because it admits solutions 
with spatially varying Coriolis fields, which do not correspond to a Newtonian field as seen in an accelerated and
rotating frame. Such non-Newtonian solutions can be excluded by imposing
appropriate boundary conditions at infinity, and will not be considered here.

\subsection{Uniform rotation\label{sec:Uniform-rotation-classic}}

The uniformly rotating frame provides a particular solution of the
Newton-Cartan field equations \eqref{NCfieldeqns} with $\rho=0$,
given by 
\begin{equation}
\begin{cases}
{\bf G}=\Omega^{2}r\,{\bf e}_{r}\\
{\bf H}=2\Omega\,{\bf e}_{z}\, ,
\end{cases}\label{eq:G_H_Uniform_Classical}
\end{equation}
where $\Omega$ is the magnitude of the constant angular velocity,
$(r,\theta,z)$ are the obvious cylindrical coordinates, and ${\bf e}_{r},{\bf e}_{\theta}$,
and ${\bf e}_{z}$ are the corresponding unit vectors. The condition
for a particle to be in circular orbit, with tangential velocity $\dot{\mathbf{r}}={\bf v}=v\,{\bf e}_{\theta},$
is that the centripetal acceleration 
\begin{equation}
\ddot{\mathbf{r}}=v\,\dot{{\bf e}}_{\theta}=-\frac{v^{2}}{r}{\bf e}_{r}
\end{equation}
equals the gravitational acceleration in Eq.~\eqref{eq:AccelS'Cartesian1}:
\begin{equation}
-\frac{v^{2}}{r}{\bf e}_{r}={\bf G}+{\bf v}\times{\bf H}=\Omega^{2}r\,{\bf e}_{r}+v\,{\bf e}_{\theta}\times2\Omega\,{\bf e}_{z}=(\Omega^{2}r+2v\Omega)\,{\bf e}_{r}\ .\label{orbit}
\end{equation}
This quadratic equation for $v$ as a function of $r$ can be solved to give
\begin{equation}
v=-\Omega r\ ,
\end{equation}
which corresponds to the particle being at rest in the non-rotating
frame. From Eq.~\eqref{orbit} it is clear that it is the Coriolis
force that provides commander Bowman's centripetal acceleration that acts against
the centrifugal force, as depicted in Fig.~\ref{Costa_NatarioFig01}(b).

We can now also understand what happens if there is an atmosphere
inside the wheel. Initially, the air will also be moving with respect
to the wheel, but friction with the walls will ultimately cause it
to come to rest. Therefore, commander Bowman will end up moving with
respect to the atmosphere, making her speed decrease due to friction. The
Coriolis force will therefore decrease and the centrifugal force will dominate,
causing her to fall ``down.'' In other words, her orbit will decay.

\subsection{The G\"odel universe\label{sec:The-Godel-universe-classic}}

The G\"odel universe is a solution of the Einstein field equations,
describing a homogeneous universe filled with a pressureless fluid of
constant vorticity.\cite{Godel,Godel2,OszvathSchucking2001,GodelTrip}
It exhibits puzzling features, such as closed timelike curves, and
the fact that it rotates rigidly about any of its points. The Newtonian
analogue provides an interesting solution of the Newton-Cartan
field equations \eqref{NCfieldeqns}, given, in a suitable frame,
by\cite{E97}
\begin{equation}
\begin{cases}
{\bf G}={\bf 0}\\
{\bf H}=2\Omega\,{\bf e}_{z}\, ,
\end{cases}\label{Godel1}
\end{equation}
where
\begin{equation}
\Omega^{2}=2\pi G\rho
\end{equation}
is a constant.  The matter fluid, of constant density $\rho$, is assumed to be at rest in this frame
(which is consistent with ${\bf G}={\bf 0}$). Notice that observers
at rest are free-falling, but measure nonzero Coriolis forces, i.e., they will see moving test particles
being deflected by a Coriolis force $m\ddot{\mathbf{r}}=m\mathbf{v}\times\mathbf{H}$.

This Newton-Cartan solution can be represented as a purely Newtonian
gravitational field by changing to a frame $S$ rotating with constant
angular velocity $-\Omega\,{\bf e}_{z}$ relative to the initial frame.
In this frame the fields become\cite{change_of_frame} 
\begin{equation}
\begin{cases}
{\bf G}=-\Omega^{2}r\,{\bf e}_{r}\\
{\bf H}={\bf 0}\; ,
\end{cases}\label{Godel2}
\end{equation}
and the interpretation of the G\"odel universe is simple: the uniform
distribution of matter, seen as an infinite cylinder, generates a
radial gravitational field ${\bf G}$ with a strength proportional to
the distance $r$ from the $z$-axis; the free-falling matter particles
(at rest in the initial frame) are moving in circular orbits with
the same angular velocity $\Omega$ in the new frame (i.e.\ with centripetal
acceleration $\ddot{\mathbf{r}}=\mathbf{G}$), so that they are rigidly
rotating around the $z$-axis. Thus, one can say that the G\"odel universe
is rotating around the $z$-axis with constant angular velocity $\Omega$. And
since the position of the $z$-axis is arbitrary (we were free to
choose the origin of the initial frame), G\"odel's universe is actually
rotating around any of its points (much like the Newtonian analogue
of the Friedmann-Lema\^\i tre-Robertson-Walker solutions can be thought
to be expanding about any of its points).\cite{M34,MM34}

The rotation around any point (which is a feature of any infinitely
wide rigidly rotating cylinder) can also be understood as follows.
Let $\mathbf{x}(t)$ be the position vector of an arbitrary fluid particle
in $S$. Its equation of motion is 
\begin{equation}
\dot{\mathbf{x}}={\bf \Omega}\times\mathbf{x}\ ,\label{eq:GodelS}
\end{equation}
where ${\bf \Omega}=\Omega\,{\bf e}_{z}$.
Now take a particular fluid particle at position $\mathbf{x}_{0}(t)$,
and consider the reference frame $S'$ with origin at $\mathbf{x}_{0}(t)$
that is not rotating with respect to $S$. The position vector of
an \emph{arbitrary} fluid particle in $S'$ is, from Eq.~(\ref{eq:x-r}),
$\mathbf{r}=\mathbf{x}-\mathbf{x}_{0}$; hence its equation of motion
is 
\begin{equation}
\dot{\mathbf{r}}=\dot{\mathbf{x}}-\dot{\mathbf{x}}_{0}={\bf \Omega}\times(\mathbf{x}-\mathbf{x}_{0})={\bf \Omega}\times\mathbf{r}\ ,\label{eq:GodelS'}
\end{equation}
which is formally identical to Eq.~(\ref{eq:GodelS})
(with $\mathbf{x}\leftrightarrow\mathbf{r}$). That is, in the frame $S'$ the fluid
is seen to be rigidly rotating about the new origin $\mathbf{r}=0$
(i.e., $\mathbf{x}=\mathbf{x}_{0}$ in the coordinates of $S$). Since
the cylinder is infinite, the picture in frame $S'$ is indistinguishable
from the picture in frame $S$. Therefore, we see that through any
point $\mathbf{x}_{0}(t)$, rotating rigidly (from the point of view
of $S$) with an angular velocity $\Omega$ about the $z$-axis,
passes an axis of rotation for the fluid indistinguishable from the
original one.

Note that this symmetry in the choice of the rotation axis is not
apparent in the purely Newtonian form \eqref{Godel2} of the fields.
In the Newton-Cartan form \eqref{Godel1}, however, it is clear that
the Newtonian analogue of the G\"odel universe is spatially homogeneous.

\subsection{Torque on a gyroscope\label{sec:Torque}}

For a rotating frame the Newton-Cartan Coriolis field is twice the
angular velocity, and so a fixed direction in an inertial frame (say,
the axis of a gyroscope) will precess with angular velocity $-{\bf \Omega}=-{\bf H}/2$.
This can also be seen starting with the equation of motion \eqref{eq:AccelS'Cartesian}
and computing the torque on a gyroscope: 
\begin{equation}
{\boldsymbol{\tau}}=\int_{\text{gyro}}{\bf r}\times(\rho{\bf G}+{\bf j}\times{\bf H}+\boldsymbol{\mathcal{F}})\,d^{3}r\ ,
\end{equation}
where ${\bf r}$ is the position vector with respect to the gyroscope's
center of mass, $\rho$ is the gyroscope's mass density, ${\bf j}=\rho{\bf v}$
is the mass current density due to the gyroscope's spinning motion,
and $\boldsymbol{\mathcal{F}}$ the force \emph{density} (arising
from the internal stresses that act on the mass elements, binding
them together). The contribution of $\boldsymbol{\mathcal{F}}$ to
${\boldsymbol{\tau}}$ is zero since the net torque of internal forces
is zero.  Thus, assuming that the gyroscope is small enough so that the fields
are constant along it, we obtain 
\begin{equation}
{\boldsymbol{\tau}}=\left(\int_{\text{gyro}}\rho{\bf r}\,d^{3}r\right)\times{\bf G}+\int_{\text{gyro}}{\bf r}\times({\bf j}\times{\bf H})\,d^{3}r\ .
\end{equation}
The first integral vanishes by definition of center of mass, and the second can be obtained
from the standard computation (see, e.g., Eq.~(5.70) in Ref.~\onlinecite{Jackson})
of the torque exerted by a magnetic field ${\bf B}$ on a magnetic
dipole ${\boldsymbol{\mu}}$ (replace the charge current and ${\bf B}$
with the mass current and ${\bf H}$). The torque is then found to be
\begin{equation}
{\boldsymbol{\tau}}=\frac{1}{2}{\bf S}\times{\bf H}\ ,\label{torque1}
\end{equation}
where ${\bf S}$ is the gyroscope's angular momentum.  Therefore, we see
that ${\bf S}$ varies as 
\begin{equation}
\frac{d{\bf S}}{dt}=-\frac{1}{2}{\bf H}\times{\bf S}\, ;\label{torque}
\end{equation}
that is, it precesses with angular velocity $-{\bf H}/2$.

The torque in Eq.~\eqref{torque1} can be interpreted as minus the derivative
with respect to $\theta$ of the ``potential energy'' $U=-{\bf S}\cdot{\bf H}/2=-\frac{1}{2}SH\cos\theta$,
in analogy with the ``potential energy''\cite{potential} $U=-{\boldsymbol{\mu}}\cdot{\bf B}$
of a magnetic dipole in a magnetic field. This ``potential energy'' explains why a {\em
gyrocompass} (which is essentially just a gyroscope\cite{Levi})
tends to align itself with the Earth's rotation axis: dissipation
effects will drive the system towards the minimum at $\theta=0$.
Note the similarity with the mechanism that makes a regular compass
align itself with the Earth's magnetic field.

\section{Inertial forces in General Relativity\label{sec:Inertial-forces-in_GR}}

The similarity between Eqs.~\eqref{NCfieldeqns} and Maxwell's equations
is suggestive, but there are important differences: the nonlinear
term ${\bf H}^{2}/2$ in the third equation, plus the fact that the
second equation has no sources. 
By analogy with electromagnetism, where magnetic effects are typically
relativistic effects sourced by charge currents (that is, of order
$v/c$, where $v$ is the speed of the charges and $c$ is the speed
of light), one might hope to find a source for the Coriolis field
in general relativity. This is indeed the case.

For simplicity, we will restrict ourselves to the
linearized theory approximation\cite{analogies} (e.g., Refs.~\onlinecite{Harris1991,
Gravitation and Inertia, Carroll, Stephani}), where one considers
a metric $g_{\alpha\beta}=\eta_{\alpha\beta}+h_{\alpha\beta}$ that
differs from the flat Minkowski metric $\eta_{\alpha\beta}$ (in rectangular
coordinates) only by small perturbations, $|h_{\alpha\beta}|\ll1$.
For our purposes it is also sufficient to consider stationary
(i.e.\ time-independent) fields, and a metric of the form 
\begin{equation}
ds^{2}=-\left(1+\frac{2\phi}{c^{2}}\right)c^{2}dt^{2}+2A_{i}\,dt\,dx^{i}+\left(1-\frac{2\phi}{c^{2}}\right)\delta_{ij}\,dx^{i}\,dx^{j}\label{metric}
\end{equation}
(which assumes $h_{ij}=h_{00}\,\delta_{ij}$). This metric describes
\emph{any} stationary, isolated matter distribution, accurate to linear
order.\cite{Stephani} As usual, we use the coordinate $x^0=ct$, and the notation $\partial_{\alpha}\equiv\partial/\partial x^{\alpha}$
for partial derivatives and $\Gamma_{\beta\gamma}^{\alpha}$ for the
Christoffel symbols (given in Appendix \ref{sub:Linearized-Christoffel-Symbols}). Latin indices $i,j,\ldots$ represent spatial coordinate indices, running from 1 to 3, whereas Greek indices $\alpha, \beta, \ldots$ represent spacetime coordinate indices, running from 0 to 3. We denote the covariant derivative by $D_{\alpha}$;
for instance, for a 4-vector $X^{\alpha}$, 
\begin{equation}
D_{\beta}X^{\alpha}=\partial_{\beta}X^{\alpha}+\Gamma_{\beta\gamma}^{\alpha}X^{\gamma}\ .\label{eq:Covariant}
\end{equation}
The linearized Ricci tensor is $R_{\beta\delta}\equiv R_{\ \beta\alpha\delta}^{\alpha}=\partial_{\alpha}\Gamma_{\beta\delta}^{\alpha}-\partial_{\delta}\Gamma_{\beta\alpha}^{\alpha}$,
which, for the metric in Eq.~(\ref{metric}), reads (see Appendix~\ref{sub:Linearized-Ricci-tensor} for details)
\begin{align}
R_{00} &= -\frac{1}{c^{2}}\nabla\cdot\mathbf{G}\, ; \\
R_{i0} &= R_{0i}=\frac{1}{2c}(\nabla\times\mathbf{H})_{i}\, ; \\
R_{ij} &= -\frac{1}{c^{2}}(\nabla\cdot\mathbf{G})\delta_{ij}\, ,
\end{align}
where 
\begin{equation}
{\bf G}=-\nabla\phi\, ;\qquad{\bf H}=\nabla\times{\bf A}\, .\label{eq:GEMfields2}
\end{equation}
The source is taken to be non-relativistic so that the energy-momentum
tensor has the components $T^{00}=\rho c^{2}$, $T^{0i}\equiv cj^{i}=\rho cv_{{\rm matter}}^{i}$,
and $T^{ij}=\mathcal{O}(\rho v_{{\rm matter}}^{2})$. Here $\rho$ and
$j^{i}$ are, respectively, the mass/energy density and current density.
Thus, the linearized Einstein field equations
\begin{equation}
R_{\alpha\beta}=\frac{8\pi G}{c^{4}}\left(T_{\alpha\beta}^{\ }-\frac{1}{2}g_{\alpha\beta}^{\ }T_{\ \gamma}^{\gamma}\right) \label{eq:Einstein_field}
\end{equation}
reduce to (cf.~Appendix \ref{sub:Linearized-Einstein-equations})
\begin{align}
\nabla\cdot\mathbf{G} &= -4\pi G\rho \label{eq:GEMfield1i}\, ; \\
\nabla\times\mathbf{H} &= -\frac{16\pi G}{c^{2}}\mathbf{j}\, .\label{eq:GEMfield1ii}
\end{align}
More precisely, both the time-time and space-space components of Eq.~(\ref{eq:Einstein_field})
yield Eq.~\eqref{eq:GEMfield1i} (in the latter case we noted that
$|T^{ij}|\ll|T^{00}|$, cf. Eq.~\eqref{eq:TabIneq}, so that the contribution of $T_{ij}$
to Eq.~(\ref{eq:Einstein_field}) is negligible), and \eqref{eq:GEMfield1ii}
is the space-time (or time-space) components of Eq.~(\ref{eq:Einstein_field}).
From Eq.~(\ref{eq:GEMfields2}) we also have 
\begin{align}
\nabla\times\mathbf{G} &= {\bf 0}\, ; \label{eq:GEMfield2i} \\
\nabla\cdot\mathbf{H} &= 0\, .\label{eq:GEMfield2ii}
\end{align}
Eqs.~(\ref{eq:GEMfield1i})--(\ref{eq:GEMfield2ii}) exhibit a striking
analogy with the Maxwell equations for electromagnetostatics. From
Eq.~\eqref{eq:GEMfield1i} we see that $\mathbf{G}$ (sometimes
called the \emph{gravitoelectric} field) is the Newtonian field (and
$\phi$ the Newtonian potential) sourced by $\rho$. From Eq.~\eqref{eq:GEMfield1ii}
we see that $\mathbf{H}$ is sourced by $\mathbf{j}$, in analogy
with the Maxwell-Amp\`ere law of electromagnetism, and is usually dubbed,
in this context, the {\em gravitomagnetic} field (see, e.g., Refs.~\onlinecite{Harris1991,
Gravitation and Inertia, Carroll, paperanalogies, GEM User Manual}).

Consider now a freely falling test particle, and let $U^{\alpha}\equiv dr^{\alpha}/d\tau$
be its 4-velocity, where $\tau$ is the proper time along its geodesic
worldline $r^{\alpha}(\tau)=(ct,\mathbf{r})$; $t=r^{0}/c$ is the
coordinate time, cf.~Eq.~(\ref{metric}). The geodesic equation
is 
\begin{equation}
U^{\beta}D_{\beta}U^{\alpha}=0\ \Leftrightarrow\ \frac{d^{2}r^{\alpha}}{d\tau^{2}}+\Gamma_{\beta\gamma}^{\alpha}\frac{dr^{\beta}}{d\tau}\frac{dr^{\gamma}}{d\tau}=0\ .\label{eq:Geo0}
\end{equation}
This equation can be expressed in terms of the coordinate time $t$.
Using $d/d\tau=(dt/d\tau)d/dt$ and $r^{0}=ct$, a straightforward
computation (see Appendix \ref{sub:Geodesic-Equation} for details) leads to 
\begin{equation}
\frac{d^{2}r^{\alpha}}{dt^{2}}+\left(\Gamma_{\beta\gamma}^{\alpha}-\frac{1}{c}\Gamma_{\beta\gamma}^{0}\frac{dr^{\alpha}}{dt}\right)\frac{dr^{\beta}}{dt}\frac{dr^{\gamma}}{dt}=0\ .\label{eq:Geo}
\end{equation}
The time component vanishes trivially. For particles moving with velocity
much smaller than the speed of light, we can keep only terms up to
first order in $v/c$, where $\mathbf{v}=d\mathbf{r}/dt\equiv\dot{\mathbf{r}}$.
Under these conditions the spatial components of Eq.~(\ref{eq:Geo})
yield, for the metric (\ref{metric}), 
\begin{equation}
\frac{d^{2}r^{i}}{dt^{2}}=-c^2\Gamma_{00}^{i}-2c\Gamma_{0j}^{i}v^{j}\ \Leftrightarrow\ \ddot{\mathbf{r}}=\mathbf{G}+\dot{\mathbf{r}}\times\mathbf{H}\ .\label{eq:FGEM_GR}
\end{equation}

This equation is analogous to the Lorentz force of electromagnetism, and it
is formally identical to Eq.~(\ref{eq:AccelS'Cartesian1}). Moreover,
in the limit $c\to\infty$, Eqs.~(\ref{eq:GEMfield1i})--(\ref{eq:GEMfield2ii})
reduce to the \emph{linearized} time-independent Newton-Cartan equations
(\ref{NCfieldeqns}). The correspondence is actually stronger: it
can be shown\cite{ExactGEM} that the exact, non-linear version of
Eqs.~(\ref{eq:GEMfield1i})--(\ref{eq:GEMfield2ii}) (which is outside
the scope of this work) reduce \emph{exactly} to Eq.~(\ref{NCfieldeqns})
when $c\to\infty$.

The reason for this correspondence is that $\mathbf{G}$
and $\mathbf{H}$ are precisely the general relativistic versions
of the fields studied in Sec.~\ref{sec:Non-relativistic-inertial-forces}.
In order to see this, let $u^{\alpha}$ denote the 4-velocity of the
observers at rest in the coordinate system $S$ of Eq.~(\ref{metric});
that is, $u^{\alpha}=(u^{0},0)$ with $u^{0}=c(-g_{00})^{-1/2}\simeq c(1-\phi /c^2)$, so that
$u_{\alpha}=g_{\alpha\beta}u^{\beta}\simeq(-c-\phi/c,\mathbf{A})$.
The 4-acceleration and 4-vorticity of these observers are defined as $a^{\alpha}=u^{\beta}D_{\beta}u^{\alpha}$
and $\omega^{\alpha}=(1/2c)\epsilon^{\alpha\gamma\delta\beta}u_{\beta}D_{\gamma}u_{\delta}$,
where $\epsilon_{\alpha\beta\gamma\delta}$ is the Levi-Civita alternating
tensor (see, e.g., Ref.~\onlinecite{paperanalogies}, whose sign
conventions we follow). The time components of these quantities are
zero in $S$ ($a^{0}=\omega^{0}=0$) and their space components
read, to linear order,
\begin{equation}
a^{i}=u^{\alpha}D_{\alpha}u^{i}=\Gamma_{00}^{i}u^{0}u^{0}\simeq\Gamma_{00}^{i}c^{2}=-G^{i}\label{eq:accelS_GR}
\end{equation}
and 
\begin{align}
\omega^{i} & =\frac{1}{2c}\epsilon_{\ \ \ \beta}^{i\gamma\delta}u^{\beta}(\partial_{\gamma}u_{\delta}-\Gamma_{\gamma\delta}^{i})=\frac{1}{2c}\epsilon_{\ \ \
0}^{ijk}u^{0}\partial_{j}u_{k}\nonumber \\
 & \simeq\frac{1}{2}\epsilon^{ijk}\partial_{j}u_{k}=\frac{1}{2}\epsilon^{ijk}\partial_{j}A_{k}=\frac{1}{2}H^{i}\ .\label{eq:VorticityS_GR}
\end{align}
Thus, $\mathbf{G}$ is minus the acceleration of the rest observers
in $S$; this is analogous to the Newtonian field in Eqs.~\eqref{eq:AccelS'Cartesian}
and \eqref{G} of Sec.~\ref{sec:Non-relativistic-inertial-forces} because
this acceleration is precisely the nongravitational force per unit
mass that must be exerted so that the observers remain at rest. On
the other hand, $\mathbf{H}$ is twice the vorticity of the observers
at rest in $S$, just like the field in Eq.~(\ref{H}) is twice the
vorticity of the observers at rest in the rotating frame $S'$ of
Sec.~\ref{sec:Non-relativistic-inertial-forces}, cf.~Eq.~(\ref{eq:VorticityNewtonian})
(note moreover that $\epsilon^{\alpha\beta\gamma\delta}u_{\beta}D_{\gamma}/2$
is a $4$-dimensional curl operator\cite{GEM User Manual}). Hence,
just like its Newtonian counterpart, the field $\mathbf{H}$ in Eq.~(\ref{eq:GEMfields2})
indeed \emph{is a Coriolis field}. However, it is now sourced by mass
currents ${\bf j}=\rho{\bf v}$, hence is not uniform in general. Analogously
to the magnetic field of electromagnetism, ${\bf H}$ can be non-zero
close to mass currents, while vanishing far from them.

\begin{figure}[h!]
\includegraphics[width=1\columnwidth]{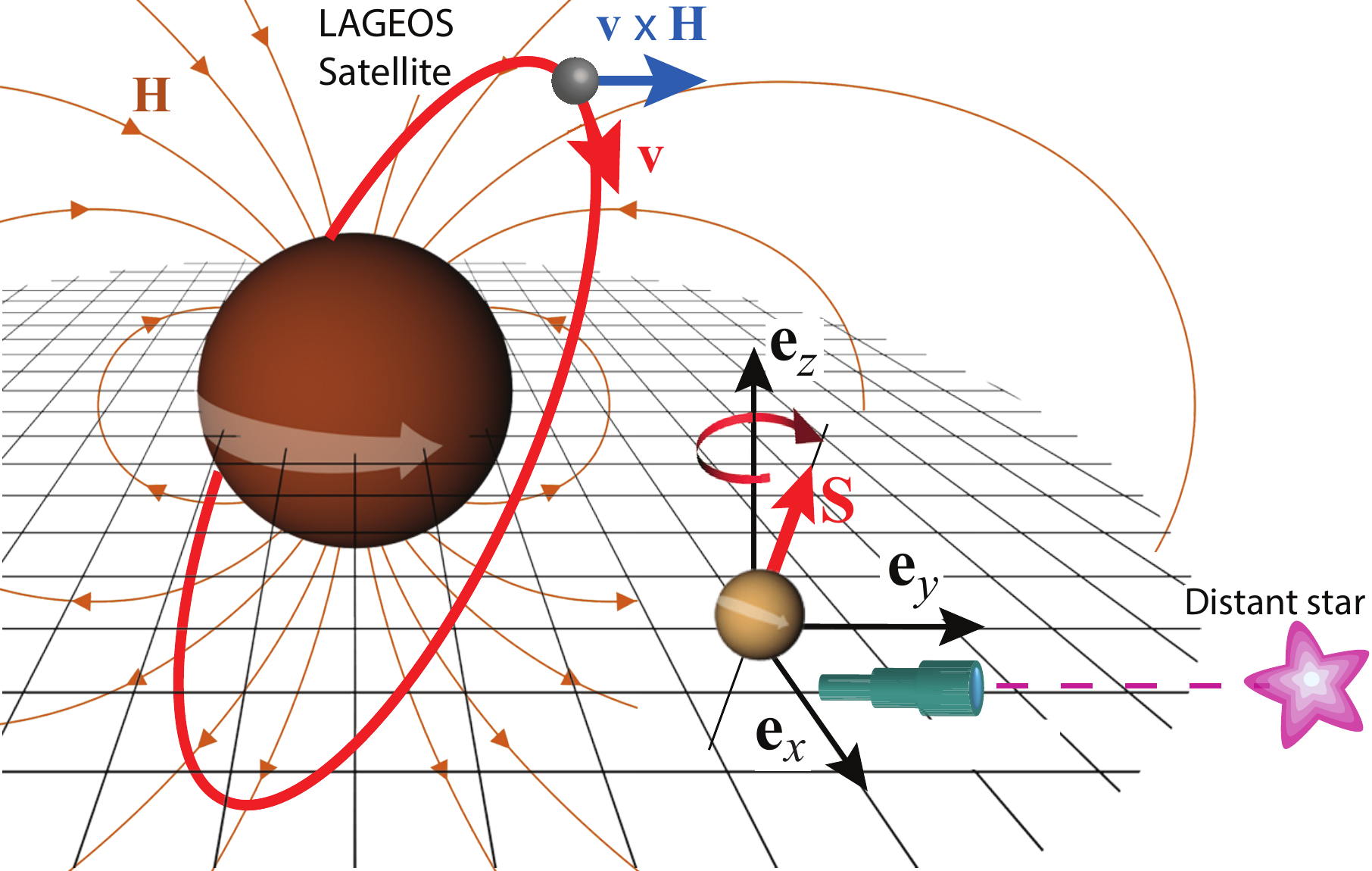}
\protect\protect\protect\caption{Star-fixed reference frame $S$: observers
at rest in this frame form a rigid grid anchored to the asymptotic inertial frame where $g_{\alpha\beta}\to\eta_{\alpha\beta}$ (assumed fixed to the distant stars; experimentally, the spatial basis
vectors $\mathbf{e}_{i}$ are set up by pointing telescopes to fixed
stars). At infinity $S$ is inertial; at finite $r$, however, it
is both accelerated and rotating. The associated Coriolis field
$\mathbf{H}$ causes gyroscopes and satellite orbits to precess relative
to $S$.}
\label{Costa_NatarioFig02}
\end{figure}

Note that $dl^{2}=g_{ij}dx^{i}dx^{j}=(1-2\phi /c^2)\delta_{ij}dx^{i}dx^{j}$
yields, to linear order (see e.g.\ Eqs.~(4.6)--(4.7) of Ref.~\onlinecite{LL97}),
the squared distance between observers at rest in $S$ as measured
by Einstein's light signaling procedure (the radar distance; see Sec.~84
of Ref.~\onlinecite{LL97} for more details). Now $dl^{2}$
is constant in time, which means these observers form a rigid
congruence (they may be thought of as fixed to a rigid grid in space,
like the one depicted in Fig.~\ref{Costa_NatarioFig02}). Because
${\bf H}$ is twice the vorticity of the observers, and the frames
we are considering are \emph{rigid}, this means that a rigid frame
$S$ can be at the same time inertial at infinity, and rotating close
to the sources, which is a manifestation of \emph{frame-dragging}.

An example, depicted in Fig.~\ref{Costa_NatarioFig02}, is the
gravitational field of a spinning celestial body, such as Earth.
In this case $\phi=-GM/r$, $\mathbf{G}=-GM\mathbf{r}/r^{3}$, and there
is also a Coriolis field \textbf{$\mathbf{H}$} created by the mass
currents due to the Earth's rotation. Given the similarity of Eq.~\eqref{eq:GEMfield1ii}
with the Maxwell-Amp\`ere law, $\mathbf{A}$ and ${\bf H}$ can be obtained
from the standard computation (e.g., Ref.~\onlinecite{Jackson},
Sec.~5.6) of the magnetic field created by a spinning charged body
(replacing the vacuum magnetic permeability $\mu_{0}$ with $-16\pi G/c^{2}$
and the charge current with the mass current), leading to $\mathbf{A}=-2G\mathbf{J}\times\mathbf{r}/(c^{2}r^{3})$
and 
\begin{equation}
{\bf H}=\frac{2G}{c^{2}}\left(\frac{{\bf J}}{r^{3}}-\frac{3({\bf J}\cdot{\bf r}){\bf r}}{r^{5}}\right)\label{eq:H-Earth}
\end{equation}
(cf., e.g., Eq.~(6.1.25) of Ref.~\onlinecite{Gravitation and Inertia}),
where ${\bf J}$ is Earth's angular momentum.

These fields are measured in the reference frame $S$ associated
to the coordinate system of the metric (\ref{metric}). The frame
$S$ is rigid and inertial \emph{at infinity} (where the metric becomes flat,
as both $\phi$ and $\mathbf{A}$ vanish); it is said to
be fixed to the ``distant stars.'' However, at \emph{finite} $r$
it is both an accelerated and rotating frame, as observers at rest
in $S$ (fixed to the rigid grid in Fig.~\ref{Costa_NatarioFig02})
are not freely falling: their acceleration is $\mathbf{a}=GM\mathbf{r}/r^{3}=-\mathbf{G}$,
as is well known, and their vorticity is $\boldsymbol{\omega}=\mathbf{H}/2$.
Note that $\boldsymbol{\omega}$ is an invariant, local measure of
rotation.  Just like in fluid dynamics, where at a given point the
vorticity yields the angular velocity of rotation of the neighboring
fluid particles (with respect to the ``compass of inertia,'' i.e.
to inertial axes), $\boldsymbol{\omega}$ yields the angular velocity
of rotation of the neighboring observers with respect to the \emph{compass
of inertia}. This so-called ``compass of inertia''\cite{Gravitation and Inertia}
is a system of \emph{absolutely} non-rotating axes, physically determined,
both in classical mechanics and in general relativity, by the spin
axes of local guiding gyroscopes. Hence, an immediate physical manifestation
of the rotation of $S$ is the fact that gyroscopes are seen to ``precess''
relative to it. %
Such precession, known as the \emph{Lense-Thirring }precession, was
recently detected by the Gravity Probe B experiment.\cite{GravityProbeB}
This precession can be obtained by the same procedure of Sec.~\ref{sec:Non-relativistic-inertial-forces},
yielding Eq.~\eqref{torque} [now in terms of the Coriolis field
(\ref{eq:H-Earth})]. Equation~\eqref{torque} actually holds even in
the exact theory, and not just in the weak field regime (cf., e.g., Eq.~(63)
of Ref.~\onlinecite{paperanalogies}).

Another manifestation of the Coriolis field \eqref{eq:H-Earth} that
arises in $S$ is the deflection of test particles due to the Coriolis
force $m\mathbf{v}\times\mathbf{H}$; it has been detected by measuring
the \emph{orbital} precession of the LAGEOS satellites,\cite{Ciufolini Lageos}
and there is an ongoing space mission (the LARES mission)\cite{LARES}
whose primary goal is to improve the accuracy of this measurement.
The effect is again readily computed in the framework above: from
Eq.~\eqref{eq:FGEM_GR} it follows that the variation of the orbital
angular momentum ${\bf L}=m{\bf r}\times{\bf v}$ is given by
\begin{equation}
\frac{d{\bf L}}{dt}=m{\bf r}\times({\bf v}\times{\bf H})\ \label{eq:LAGEOS}
\end{equation}
(since ${\bf r}\times{\bf G}={\bf 0}$).

We close this section with the following remarks. We have seen that
$\mathbf{H}$ is a Coriolis field formally similar to the one of Newtonian
mechanics (and as such, a field of \emph{fictitious} forces). But
it is purely relativistic, in the sense that it arises in a
\emph{star fixed frame} (where its Newtonian counterpart vanishes)
due to the frame-dragging effect produced by mass currents. It is instructive
to think about frame-dragging in terms of a Coriolis field, as it
averts some common misconceptions about this effect in the literature
(in particular those arising from the fluid dragging analogy proposed
by some authors); for a detailed discussion of these problems we refer
the reader to Ref.~\onlinecite{Rindler}.

\section{Conclusion\label{sec:conclusion}}

In this work we discussed the Coriolis field in different settings;
although its effects are familiar in some situations, there exist other, seemingly
mysterious phenomena, that are simply effects of a Coriolis field.

We started with the example of an astronaut freely falling inside
a rotating ship, whose motion from the point of view of the ship frame
is puzzling until one realizes the role of the Coriolis field. We
proceeded by showing that the Coriolis field satisfies the field equations
of the so-called Newton-Cartan theory, a generalization
of Newtonian theory that is covariant under changes of rigid (arbitrarily
accelerated and rotating) frame. We presented simple solutions of this theory,
including the Newtonian analogue of the G\"odel universe, whose homogeneous
rotation is easily understood in this framework. 

Finally, we discussed
the purely relativistic Coriolis field generated by mass currents.
It is usually cast in the literature as a ``gravitomagnetic field,''
whose physical meaning---namely that it is
just a Coriolis field, arising from the fact that the star fixed reference
frame is in fact rotating close to mass currents---is often not transparent. 
The Lense-Thirring precession of the Gravity Probe B mission gyroscopes is
seen to come from the same principle as the Newtonian precession of a
gyrocompass relative to an Earth-fixed frame, and both are obtained by a
computation analogous to that yielding the precession of a magnetic dipole
in a magnetic field.
The Lense-Thirring orbital precession of the LAGEOS/LARES satellites
is also simply explained in terms of the Coriolis force, analogous
to the Newtonian deflection of test particles in a rotating frame.
Thinking in terms of a Coriolis field gives a correct, simple interpretation
of the frame-dragging effect, avoiding the misconceptions about this
effect that are all too common in the literature.

\begin{acknowledgments}
We thank the referees for the useful comments and suggestions that helped us
improve this paper, and Rui Quaresma for helping us with the illustrations. This work was partially funded
by FCT/Portugal through project PEst-OE/EEI/LA0009/2013. L.~F.~C.~is
funded by FCT through grant SFRH/BDP/85664/2012.
\end{acknowledgments}

\appendix

\section{Formulae for Sec.~\ref{sec:Inertial-forces-in_GR}\label{sec:Formulae-for-Sec.}}

\subsection{Linearized Christoffel Symbols\label{sub:Linearized-Christoffel-Symbols}}

The Christoffel symbols are given, in terms of the metric tensor,
as (e.g., Eq.~(3.1) of Ref.~\onlinecite{Carroll})
\begin{equation}
\Gamma_{\beta\gamma}^{\alpha}=\frac{1}{2}g^{\alpha\delta}(\partial_{\gamma}g_{\delta\beta}+\partial_{\beta}g_{\delta\gamma}-\partial_{\delta}g_{\beta\gamma})\ .
\end{equation}
For the metric in Eq.~\eqref{metric}, the \emph{linearized} Christoffel
symbols are obtained by neglecting all the terms that are not linear in the
perturbations $\phi$ and $A_{i}$; they read 
\begin{align}
\Gamma_{00}^{0}&=0\ ;\quad\Gamma_{00}^{i}=\frac{1}{c^{2}}\partial^{i}\phi=-\frac{1}{c^{2}}G^{i}=\Gamma_{i0}^{0}\ ;\\
\Gamma_{jk}^{0}&=-\frac{1}{2c}(\partial_{j}A_{k}+\partial_{k}A_{j})\ ;\quad\Gamma_{0k}^{j}=\frac{1}{2c}(\partial_{k}A^{j}-\partial^{j}A_{k})\ ;\\
\Gamma_{kl}^{j}&=\frac{1}{c^{2}}(\delta_{kl}\partial^{j}\phi-\delta^{j}_{l}\partial_{k}\phi-\delta^{j}_{k}\partial_{l}\phi) \nonumber \\
&=\frac{1}{c^{2}}(\delta^{j}_{l}G_{k}+\delta^{j}_{k}G_{l}-\delta_{kl}G^{j})\ ,
\end{align}
where we used the fact that, since $\phi$ and $A_{i}$ are time independent,
all the time derivatives $\partial_{0}=(1/c)\partial/\partial t$
vanish. Note also that, to linear order, spatial indices are raised and
lowered with the Kronecker delta, and so their vertical position is immaterial.

\subsection{Linearized Ricci tensor\label{sub:Linearized-Ricci-tensor}}

The Ricci tensor $R_{\beta\delta}$ is defined in terms of the Riemann
curvature tensor $R_{\ \beta\gamma\delta}^{\alpha}$ as $R_{\beta\delta}\equiv R_{\ \beta\alpha\delta}^{\alpha}$.
The Riemann tensor is given by, e.g., Eq.~(3.113) of Ref.~\onlinecite{Carroll};
to linear order $R_{\ \beta\gamma\delta}^{\alpha}=\partial_{\gamma}\Gamma_{\beta\delta}^{\alpha}-\partial_{\delta}\Gamma_{\beta\gamma}^{\alpha}$,
and so
\begin{equation}
R_{\beta\delta}=\partial_{\alpha}\Gamma_{\beta\delta}^{\alpha}-\partial_{\delta}\Gamma_{\beta\alpha}^{\alpha}\ .
\end{equation}
For the metric \eqref{metric}, the time-time component reads
\begin{equation}
R_{00}=\partial_{k}\Gamma_{00}^{k}=\frac{1}{c^{2}}\partial^{k}\partial_{k}\phi=-\frac{1}{c^{2}}\nabla\cdot\mathbf{G}\, ,\label{eq:R00}
\end{equation}
while the time-space components $R_{0i}=R_{i0}$ are
\begin{equation}
R_{0i}=\partial_{j}\Gamma_{0i}^{j}=\frac{1}{2c}\partial^{j}\left(\partial_{i}A_{j}-\partial_{j}A_{i}\right)\, .
\end{equation}
Noting that $H^{k}=(\nabla\times\textbf{A})^{k}=\epsilon^{kmn}\partial_{m}A_{n}$,
it follows that $\epsilon_{ijk}H^{k}=\epsilon_{kij}\epsilon^{kmn}\partial_{m}A_{n}=\partial_{i}A_{j}-\partial_{j}A_{i}$,
and therefore
\begin{equation}
R_{0i}=\frac{1}{2c}\epsilon_{ijk}\partial^{j}H^{k}=\frac{1}{2c}(\nabla\times \textbf{H})_{i}\ .\label{eq:Ri0}
\end{equation}
Lastly, the space-space components are
\begin{align}
R_{ij} & =\partial_{k}\Gamma_{ij}^{k}-\partial_{j}\Gamma_{ik}^{k}-\partial_{j}\Gamma_{i0}^{0}=-\frac{1}{c^{2}}(\nabla\cdot\mathbf{G})\delta_{ij}\ . \label{eq:Rij}
\end{align}

\subsection{Linearized Einstein equations\label{sub:Linearized-Einstein-equations}}

For a non-relativistic source, $T^{00}=\rho c^{2}$, $T^{0i}\equiv cj^{i}=\rho cv_{{\rm matter}}^{i}$,
$T^{ij}=\mathcal{O}(\rho v_{{\rm matter}}^{2})$; this means that
\begin{equation}
|T^{ij}|\sim|T^{00}|\frac{v_{{\rm matter}}^{2}}{c^{2}}\ll|T^{00}|\ , \label{eq:TabIneq}
\end{equation}
and therefore $T_{\ \gamma}^{\gamma}=T_{\ 0}^{0}+T_{\ i}^{i}\simeq T_{\ 0}^{0}=-\rho c^{2}$.
The Einstein equations \eqref{eq:Einstein_field} then become, in
this regime,
\begin{equation}
R_{\alpha\beta}=\frac{8\pi G}{c^{4}}\left(T_{\alpha\beta}^{\ }+\frac{1}{2}g_{\alpha\beta}^{\ }\rho c^{2}\right)\ .\label{eq:Einstein2}
\end{equation}
To linear order (which implies neglecting also terms involving products
of $\rho$ with the metric perturbations), the time-time component
of Eq.~\eqref{eq:Einstein2} yields $R_{00}=4\pi G\rho/c^{2}$; equating
this expression to Eq.~\eqref{eq:R00} leads to Eq.~\eqref{eq:GEMfield1i}.
The time-space components yield
\begin{equation}
R_{0i}=\frac{8\pi G}{c^{3}}T_{0i}=-\frac{8\pi G}{c^{3}}j_{i}\, ,
\end{equation}
and equating this expression to Eq.~\eqref{eq:Ri0} yields Eq.~\eqref{eq:GEMfield1ii}.
As for the space-space components, using \eqref{eq:TabIneq} we get
$R_{ij}=4\pi G\rho\delta_{ij}/c^{2}$, and equating to Eq.~\eqref{eq:Rij}
leads to Eq.~\eqref{eq:GEMfield1i} (the same as the time-time component).

\subsection{Geodesic Equation\label{sub:Geodesic-Equation}}

Starting from the geodesic equation \eqref{eq:Geo0}, by repeated
use of $d/d\tau=(dt/d\tau)d/dt$, one obtains
\begin{align}
& \frac{d}{d\tau}\left(\frac{dt}{d\tau}\frac{dr^{\alpha}}{dt}\right)+\left(\frac{dt}{d\tau}\right)^{2}\Gamma_{\beta\gamma}^{\alpha}\frac{dr^{\beta}}{dt}\frac{dr^{\gamma}}{dt} =0\nonumber \\
& \Leftrightarrow\frac{dr^{\alpha}}{dt}\frac{d^{2}t}{d\tau^{2}}+\left(\frac{dt}{d\tau}\right)^{2}\frac{d^{2}r^{\alpha}}{dt^{2}}+\left(\frac{dt}{d\tau}\right)^{2}\Gamma_{\beta\gamma}^{\alpha}\frac{dr^{\beta}}{dt}\frac{dr^{\gamma}}{dt} =0\ . \label{eq:GeoStep1}
\end{align}
Now, since $t=r^{0}/c$, using Eq.~\eqref{eq:Geo0}, we have
\begin{align}
\frac{d^{2}t}{d\tau^{2}} & =\frac{1}{c}\frac{d^{2}r^{0}}{d\tau^{2}}=-\frac{1}{c}\Gamma_{\beta\gamma}^{0}\frac{dr^{\beta}}{d\tau}\frac{dr^{\gamma}}{d\tau} \nonumber \\
 & =-\frac{1}{c}\left(\frac{dt}{d\tau}\right)^{2}\Gamma_{\beta\gamma}^{0}\frac{dr^{\beta}}{dt}\frac{dr^{\gamma}}{dt}\, .
\end{align}
Substituting this expression back into Eq.~\eqref{eq:GeoStep1} leads to Eq.~\eqref{eq:Geo}.


\begin{thebibliography}{}
\bibitem{change_of_frame} This is best seen by considering the inverse
change of frame, to which Eqs.~\eqref{G} and \eqref{H} apply.
\bibitem{analogies} There is a huge literature about the analogies
between electromagnetism and gravity in different regimes; see Refs.~\onlinecite{paperanalogies,
GEM User Manual} and references therein.
\bibitem{ExactGEM}The exact version of Eqs.~(\ref{eq:GEMfield1i})--(\ref{eq:GEMfield2ii})
is obtained from Eqs.~(90)--(91) and~(93)--(94) of Ref.~\onlinecite{paperanalogies}
taking the case of rigid frames ($K_{(ij)}=\theta=0$ therein) and
a dust source, and restoring the $c$ and $G$ factors ($c=G=1$ in
the unit system of Ref.~\onlinecite{paperanalogies}). They read
$\tilde{\nabla}\cdot\mathbf{G}=-4\pi G(2\gamma^{2}-1)\rho+\mathbf{H}^{2}/2+\mathbf{G}^{2}/c^2$,
$\tilde{\nabla}\times\mathbf{H}=-(16\pi G/c^{2})\mathbf{j}+(2/c^{2})\mathbf{G}\times\mathbf{H}$,
$\tilde{\nabla}\times\mathbf{G}=-\dot{\mathbf{H}}$, $\tilde{\nabla}\cdot\mathbf{H}=-\mathbf{G}\cdot\mathbf{H}/c^2$,
where $\gamma=(1-v_{{\rm matter}}^{2}/c^{2})^{-1/2}$. These equations
yield Eqs.~(\ref{NCfieldeqns}) exactly when $c\rightarrow\infty$.
\end{thebibliography}
\end{document}